\documentclass[%
reprint,
 amsmath,amssymb,
 aps,
]{revtex4-1}

\usepackage{graphicx}% Include figure files
\usepackage{dcolumn}% Align table columns on decimal point
\usepackage{bm}% bold math
\usepackage[mathlines]{lineno}% Enable numbering of text and display math
\usepackage[colorlinks=true,citecolor=blue,linkcolor=blue,anchorcolor=green, anchorcolor=green,urlcolor=blue]{hyperref}
\usepackage{mathrsfs}
\usepackage{xcolor}
\usepackage{float}
\usepackage{ulem}
\usepackage[utf8]{inputenc}
\begin{document}

%\preprint{APS/123-QED}

\title{Photon orbits and phase transitions in Born-Infeld-dilaton black holes}% Force line breaks with \\

\author{Haitang Li}
\email{lhaitang@zjut.edu.cn}
\author{Yong Chen}
\email{chenyong@zjut.edu.cn}
\author{Shao-Jun Zhang}
\email{sjzhang84@hotmail.com}
\affiliation{Institute for Theoretical Physics $\&$ Cosmology, Zhejiang University of Technology, Hangzhou 310023, China}

\date{\today}% It is always \today, today,

\begin{abstract}

Relations between photon orbits and thermodynamical phase transitions are explored in Born-Infeld-dilaton-AdS black hole. The coupling between the electromagnetic field and the dialton field is chosen such that the full phase diagram contains zeroth-order and first-order phase transitions as well as RPT. We find that there exist non-monotonic beahviors of the photon orbit radius $r_{ps}$ and the minimum impact parameter $u_{ps}$ which signal the existence of the various phase transitions. In particular, the marginal value of pressure under which RPT occur can be read off from these behaviors. Along the co-existing lines, there are changes of both $r_{ps}$ and $u_{ps}$, whose dependence on the transition temperature show characteristic behaviors signalling the existence of RPT. Moreover, the critical exponents of $\Delta r_{ps}$ and $\Delta u_{ps}$ are found to take a universal value $\frac{1}{2}$. These results imply that $r_{ps}$ and $u_{ps}$ can be used as order parameters to describe BH phase transitions.
%\begin{description}
%04.70.Dy, 04.70.-s
%\end{description}
\end{abstract}

%\pacs{04.70.Dy, 04.70.-s}% PACS, the Physics and Astronomy

\maketitle

%\tableofcontents

\section{Introduction}\label{partone}

In the past decades, black hole (BH) thermodynamics has attracted lots of attention~\cite{Bekenstein:1973ur,Hawking:1974sw}. By combing gravity and quantum mechanics semi-classically, it provides us new insights into the yet unsolved quantum gravity problem as well as indicating a possible deep connection between black hole physics and other areas of fields via the AdS/CFT~\cite{Maldacena:1997re,Gubser:1998bc,Witten:1998qj}. It is found that not only do BHs have Hawking temperature and Hawking-Bekenstein entropy, they also posses rich phase structures analogous to ordinary thermodynamic systems \cite{Bardeen:1973gs,Hawking:1982dh}. In the extended phase space where the negative cosmological constant is considered as thermodynamic pressure, BHs exhibit phase transitions which in may respects similar to various chemical phenomena of ordinary thermodynamic systems, such as solid-liquid phase transition~\cite{Kubiznak:2014zwa}, liquid-gas transition of van der Waals fluid~\cite{Chamblin:1999tk,Chamblin:1999hg,Dolan:2011xt,Kubiznak:2012wp}, triple points \cite{Altamirano:2013uqa}, reentrant phase transition (RPT)~\cite{Gunasekaran:2012dq,Altamirano:2013ane}, heat engines \cite{Johnson:2014yja} and etc, and thus is sometimes referred as BH chemistry~\cite{Kastor:2009wy,Kubiznak:2014zwa,Mann:2016trh}. For more details, please refer to a most recent review~\cite{Kubiznak:2016qmn} and references therein.

Although experimental evidences of the BH thermodynamics have not been found yet, it is expected that BH phase transitions should have some observational signatures, such as quasinormal modes (QNMs), which may be detected from astronomical observations in the future. In fact, it is interesting to find that QNMs indeed will exhibit characteristic behaviors along the BH phase transitions and thus can be used as a dynamical probe to investigate the transitions \cite{Jing:2008an,Berti:2008xu,He:2008im,Koutsoumbas:2006xj,Koutsoumbas:2008pw,Shen:2007xk,Rao:2007zzb,Myung:2008ze,Liu:2014gvf,Mahapatra:2016dae,Chabab:2016cem,Prasia:2016esx,Zou:2017juz,Liang:2017ceh,Li:2017kkj}. For example, in Ref. \cite{Liu:2014gvf}, the authors pointed out that there is a dramatic change of the slope of QNMs along with the BH phase transition. These results imply a close connection between the dynamical and thermodynamical aspects of BHs.

On the other hand, it has been known for many years that QNMs are intimately linked to the existence of unstable photon orbits~\cite{Goebel:1972,Ferrari:1984zz,Mashhoon:1985cya}, with the latter playing an important role in strong gravitational phenomena, such as the shadow, lensing as well as gravitational waves \cite{Bozza:2002zj,Cardoso:2008bp,Stefanov:2010xz,Wei:2011zw,Hod:2012nk,Wei:2013mda,Raffaelli:2014ola,Cardoso:2016rao,Konoplya:2017wot}. Roughly, QNMs can be seen as the vibration frequencies of the photon orbits and thus can be derived using the properties of the photon orbits \cite{Dolan:2009nk}. Based on these observations, it is then natural to link the photon orbits and BH phase transitions and to see if properties of photon orbits can signal the BH phase transitions. 

Early in Refs. \cite{Cvetic:2016bxi,Gibbons:2008ru},, the authors pointed out that the presence of photon orbits signals a possible York-Hawking-Page type phase transition. Recently, in Ref. \cite{Wei:2017mwc}, the authors make the relations more clearly. In this work, $d$-dimensional Reissner-Nordstrom-AdS (RNAdS) BH is considered where there exists a first-order van der Waals (vdW)-like phase transition between large black hole (LBH) and small black hole (SBH) for the pressure $P$ under the critical value $P_c$. By studying the behaviors of the photon orbit radius $r_{ps}$ and the minimum impact parameter $u_{ps}$ along with the temperature, they found that the two quantities both exhibit non-monotonic behaviors for $P<P_c$ signalling the existence of phase transition. Moreover, along with the phase transition, there are changes of both $r_{ps}$ and  $u_{ps}$ which vanish at the critical point, thus implying that the two quantities can serve as order parameters for the vdW-like phase transition. In particular, these changes have an universal exponent $\frac{1}{2}$ near the critical point. Later, the study is extended to other black holes and gravity theories, for example rotating case \cite{Wei:2018aqm}, Gauss-Bonnet case \cite{Han:2018ooi}, massive gravity case \cite{Chabab:2019kfs} and Born-Infeld case \cite{Xu:2019yub}, which confirm the relations further. See also Ref. \cite{Zhang:2019tzi} for a general proof of the existence of these relations.

In this work, we would like to extend this study to Born-Infeld-dilaton-AdS (BIDAdS) black hole \cite{Dehyadegari:2017flm,Momennia:2017hsc,Dayyani:2017fuz,Sheykhi:2006dz,Sheykhi:2007gw}. The motivation to choose such kind of BH is that BIDAdS BH has rather rich phase structures which depends on the coupling constant $\alpha$ between the electromagnetic field and dilaton field. For $\alpha$ in a certain range, the full phase diagram will contain three types of phase transitions depending on the pressure: zeroth-order and first-order LBH-SBH phase transitions as well as RPT. So, it will be an ideal background to check and improve the above mentioned relations between photon orbits and phase transitions. In particular, we would like to see if the photon orbit radius and the minimum impact parameter can be used to distinguish different types of phase transitions.

The work is organized as follows. In Sec. II, phase transitions in Born-Infeld-dilaton BH are reviewed. In Sec. III, relations between photon orbits and various phase transitions are discussed. The last section is devoted to summary and discussions.

\section{PHASE TRANSITIONs in Born-Infeld-DILATON BLACK HOLE}\label{SecII}

We consider the following four-dimensional Born-Infled-dilaton black hole~\cite{Dehyadegari:2017flm,Momennia:2017hsc,Dayyani:2017fuz,Sheykhi:2006dz,Sheykhi:2007gw}
\begin{eqnarray}
	ds^2 &=& -f(r) dt^2 + f^{-1} (r) dr^2 + r^2 R^2(r) (d\theta^2 + \sin^2\theta d\phi^2),\nonumber\\
	A &=& A_t(r) dt,\quad \Phi = \Phi(r).
\end{eqnarray}
with
\begin{eqnarray}
	f(r) &=& - \frac{\alpha^2 + 1}{\alpha^2 -1} \left(\frac{b}{r}\right)^{-2\gamma} - \frac{m}{r^{1-2\gamma}}\nonumber\\
	&& + \frac{(\alpha^2 +1)^2 r^2}{\alpha^2 -3} \left(\frac{b}{r}\right)^{2\gamma} \Bigg\{\Lambda +\nonumber\\
	&&2\beta^2 \left[_2{\cal F}_1 \left(-\frac{1}{2}, \frac{\alpha^2 -3}{4}, \frac{\alpha^2 +1}{4}, -\eta\right) -1\right]\Bigg\},\nonumber\\
	R(r) &=& e^{\alpha \Phi},\nonumber\\
	A_t (r) &=& \frac{q}{r}\ _2{\cal F}_1 \left(\frac{1}{2}, \frac{\alpha^2 + 1}{4}, \frac{\alpha^2 +5}{4}, -\eta\right),\nonumber\\
	\Phi (r) &=& \frac{\gamma}{\alpha} \ln\left(\frac{b}{r}\right).
\end{eqnarray}
Here $_2{\cal F}_1$ is the hypergeometric function, and $\gamma \equiv \frac{\alpha^2}{\alpha^2 +1}, \eta \equiv \frac{q^2}{\beta^2 r^4} \left(\frac{b}{r}\right)^{4\gamma}$. The mass $M$ and charge $Q$ of the BH are,
\begin{eqnarray}
M = \frac{m b^{2 \gamma}}{2 (\alpha^2 +1)},\qquad Q = q.
\end{eqnarray}
From the metric, it is easy to obtain the Hawking temperature and Bekenstein-Hawking entropy of the BH
\begin{eqnarray}\label{Hawking-temperature}
T &=& \frac{\alpha^2 +1}{2\pi r_+ (\alpha^2 - 1)} \left(\frac{b}{r_+}\right)^{2\gamma} \bigg[r_+^2 (\alpha^2 - 1) \left(\beta^2 -\frac{\Lambda}{2}\right)\nonumber\\
&& - \frac{1}{2} \left(\frac{r_+}{b}\right)^{4\gamma} - \beta^2 r_+^2 (\alpha^2 - 1) \sqrt{\eta_+ + 1} \bigg],\nonumber\\
S &=& \pi r_+^2 \left(\frac{b}{r_+}\right)^{2\gamma},
\end{eqnarray}
where $r_+$ is the outmost horizon radius and $\eta_+ \equiv \eta\big|_{r=r_+}$.

In extended phase space, the cosmological constant is considered as the thermodynamical pressure with the precise relation in our case as
\begin{eqnarray}
P = -\frac{\Lambda}{8\pi} \left(\frac{b}{r_+}\right)^{2\gamma}.
\end{eqnarray}
As $\alpha \rightarrow 0$, it reduces to the standard one $P = -\frac{\Lambda}{8\pi}$. In terms of the Hawking temperature (\ref{Hawking-temperature}), it can be cast into a form of equation of state as
\begin{eqnarray}
P = P(r_+, T) =&& \frac{T}{2 r_+ (\alpha^2 +1)} + \frac{1}{8\pi r_+^2} \Bigg[\frac{1}{\alpha^2 -1} \left(\frac{b}{r_+}\right)^{-2\gamma}\nonumber\\
&& - 2 \beta^2 r_+^2 \left(\frac{b}{r_+}\right)^{2\gamma} \left(1 - \sqrt{\eta_+ + 1}\right)\Bigg].
\end{eqnarray}
We will work in canonical ensemble where $Q$ and $\beta$ are fixed and the Gibbs free energy is defined as
\begin{eqnarray}
G = M - T S.
\end{eqnarray}
By studying the behaviour of the Gibbs free energy versus the temperature for fixed pressure, the phase structure and transitions of the system can be depicted which have been studied thoroughly in Refs.~\cite{Dehyadegari:2017flm,Momennia:2017hsc,Dayyani:2017fuz,Chen:2018icg}. The full $P-T$ phase diagram is rather diverse depending on the set of free parameters $(q, \alpha, \beta, b)$. Following Ref. \cite{Chen:2018icg}, we fix $q=0.2, b=1$ and $\beta=2$ to study the effect of $\alpha$ on the phase diagram. It should be noted that $\alpha<1$ to ensure the positivity of the Hawking temperature. Depending on the value of $\alpha$, there  may be zeroth-order or first-order phase transition or RPT. In this work, we will take $\alpha=0.1$ when the phase diagram is richest. 

\begin{figure}[!htbp]
	
	\includegraphics[width=0.5\textwidth]{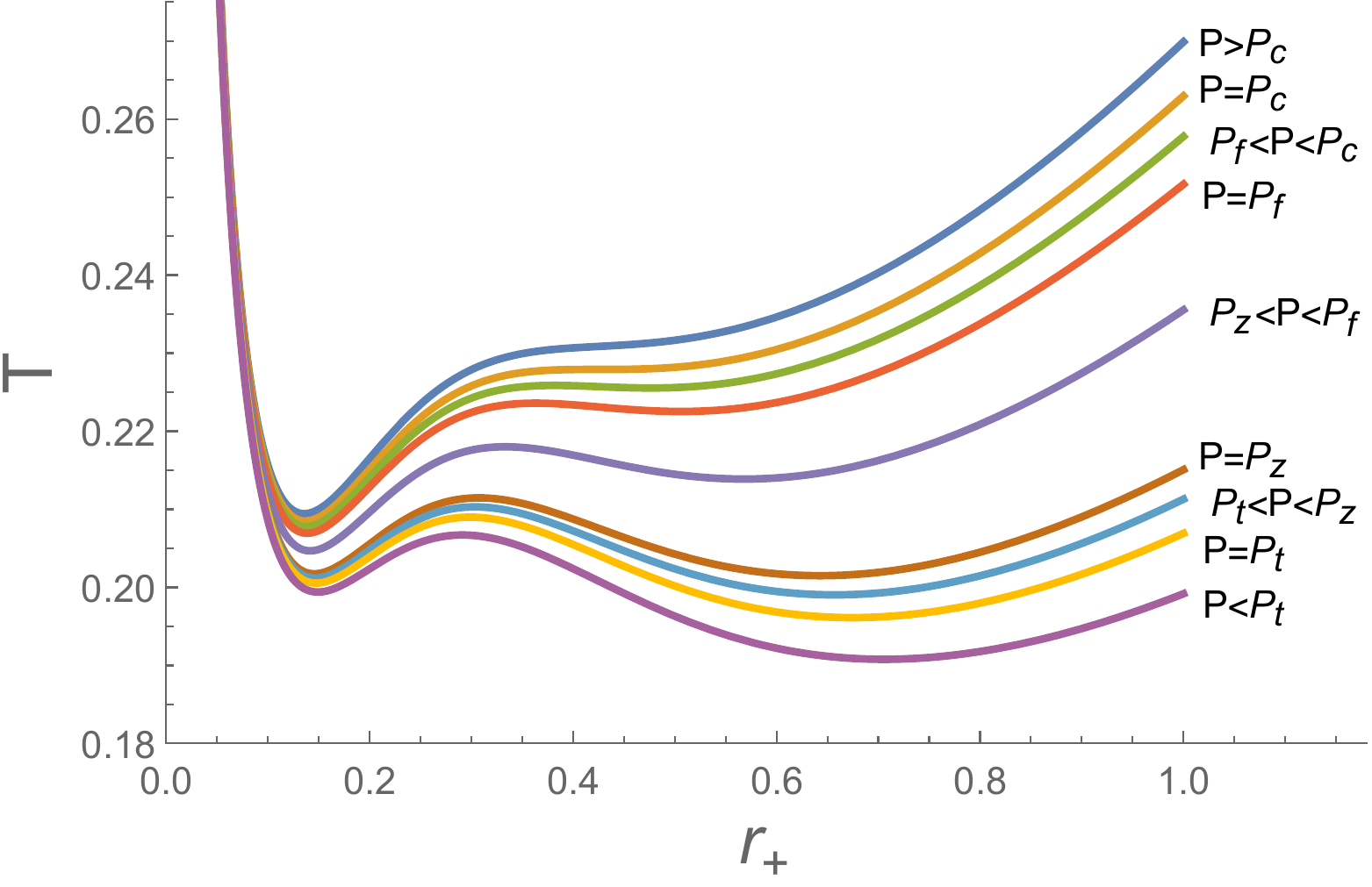}
	\caption{(colour online) Isobaric curves for various pressure.} \label{Trp}
\end{figure}

\begin{figure}[!htbp]
	\includegraphics[width=0.5\textwidth]{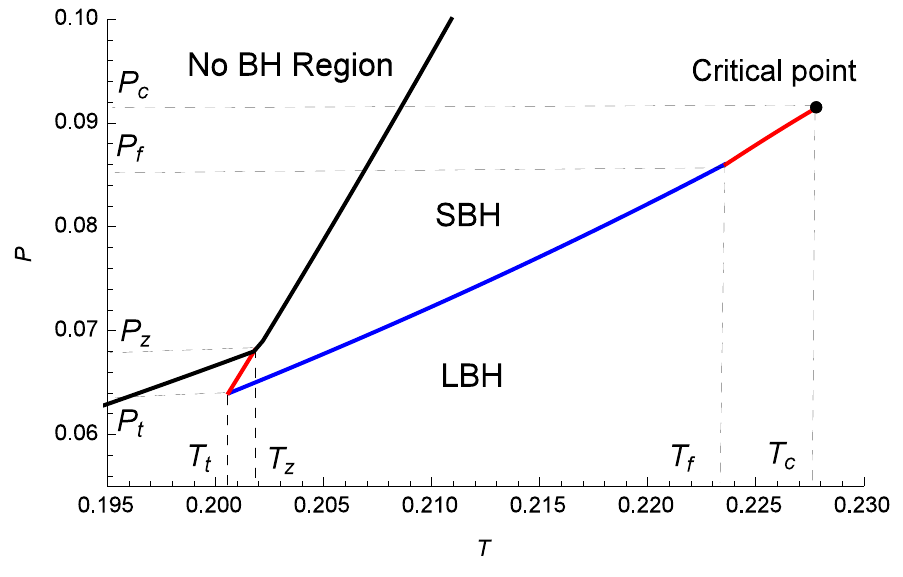}
	\caption{(colour online) $P-T$ phase diagram with $\alpha=0.1$ \cite{Chen:2018icg}. $(T_c, P_c)$ is the critical point. The red solid lines denote the co-existing line of zeroth-order SBH-LBH phase transition, while the blue solid line denotes the co-existing line of first-order LBH-SBH phase transition. The black solid lines denote the boundary for the existence of black hole solutions.}\label{PTPhaseDiagram}
\end{figure}

In Fig.~\ref{Trp}, the isobaric curves in $T-r_+$ plane are plotted for various pressure. The critical point is determined by $\partial_{r_+} P(r_+, T) \big|_{T=T_c,r_+ = r_{+c}}=\partial^2_{r_+} P(r_+, T) \big|_{T=T_c,r_+ = r_{+c}}=0$ and is $(T_c, P_c) = (0.2279,0.0916)$. From the figure, one can see that when $P<P_c$, each curve can be divided into at most four branches, out of which two are stable ( large black hole (LBH) and small black hole (SBH) both with $\frac{\partial T}{\partial r_+}>0$) while the rest are unstable ($\frac{\partial T}{\partial r_+}<0$). There may be phase transitions between LBH and SBH which can be determined by comparing their Gibbs free energy. The full phase diagram contains a zeroth-order and first-order phase transition as well as RPT, as depicted in Fig. \ref{PTPhaseDiagram}.  From the figure, one can see that there are three types of phase transitions depending on the pressure: When $P \in (P_f, P_c)$, there is a zeroth-order LBH-SBH phase transition; When $P \in (P_z, P_f)$, the transition becomes first-order; When $P \in (P_t, P_z)$, a LBH-SBH-LBH RPT appears with LBH-SBH transition being first-order and SBH-LBH zeroth-order; When $P>P_c$ or $P<P_t$, no phase transition occurs.

\section{Photon orbits and its relation with phase transitions}\label{partthree}

\subsection{Photon orbits}

In this section, we are going to discuss the relationship between photon orbits and the above mentioned thermodynamical phase transitions. Considering a free particle orbiting around the black hole on the equatorial hyperplane with $\theta=\frac{\pi}{2}$. The reduced metric on the geodesic is
\begin{equation}
ds^2=-f(r)dt^2+\frac{dr^2}{f(r)}+r^2R^{2}(r)d\phi^2.
\end{equation}
Then the Lagrangian governing the motion of the  particle is
\begin{equation}
2 \mathcal{L}=g_{\mu\nu} \dot{x}^{\mu} \dot{x}^{\nu}=-f(r)\dot{t}^2+\frac{\dot{r}^2}{f(r)}+r^2R^{2}(r)\dot{\phi}^2,
\end{equation}
where dot means derivative with respect to some affine parameter of the geodesic. For the BIDAdS BH spacetime, the geodesic admits two integral constants $E$ and $L$ associated with the two Killing vectors $\partial_t$ and $\partial_\phi$, where $E$ and $L$ are the energy and orbital angular momentum of the particle respectively. From the Lagrangian, we can obtain the canonical momentum of the particle 
\begin{eqnarray}
&&p_t=-f(r)\dot{t}=-E,\nonumber\\
&&p_r=f^{-1}(r)\dot{r},\nonumber\\
&&p_{\phi}=r^2R^2(r)\dot{\phi}=L,
\end{eqnarray}
from which one can get
\begin{eqnarray}
&&\dot{t}=\frac{E}{f(r)},\nonumber\\
&&\dot{\phi}=\frac{L}{r^2R^2(r)}
\end{eqnarray}
Using the normalization condition $g_{\mu\nu}\dot{x}^{\mu}\dot{x}^{\nu}=-\delta^2$, where $\delta^2=0$ for null geodesics and $\delta^2=1$ for timelike geodesics, we can obtain the equation of motion along the radial direction which takes a form as
\begin{equation}
\dot{r}^2 + V_{\text{eff}} = 0,
\end{equation}
where the effective potential $V_{\text{eff}}$ is
\begin{equation}\label{EffectivePotential}
V_{\text{eff}}=\frac{L^2}{r^2R^2(r)}f(r) + \delta^2 f(r)-E^2.
\end{equation}

\begin{figure}[!htbp]
  \centering
  \includegraphics[width=0.50\textwidth]{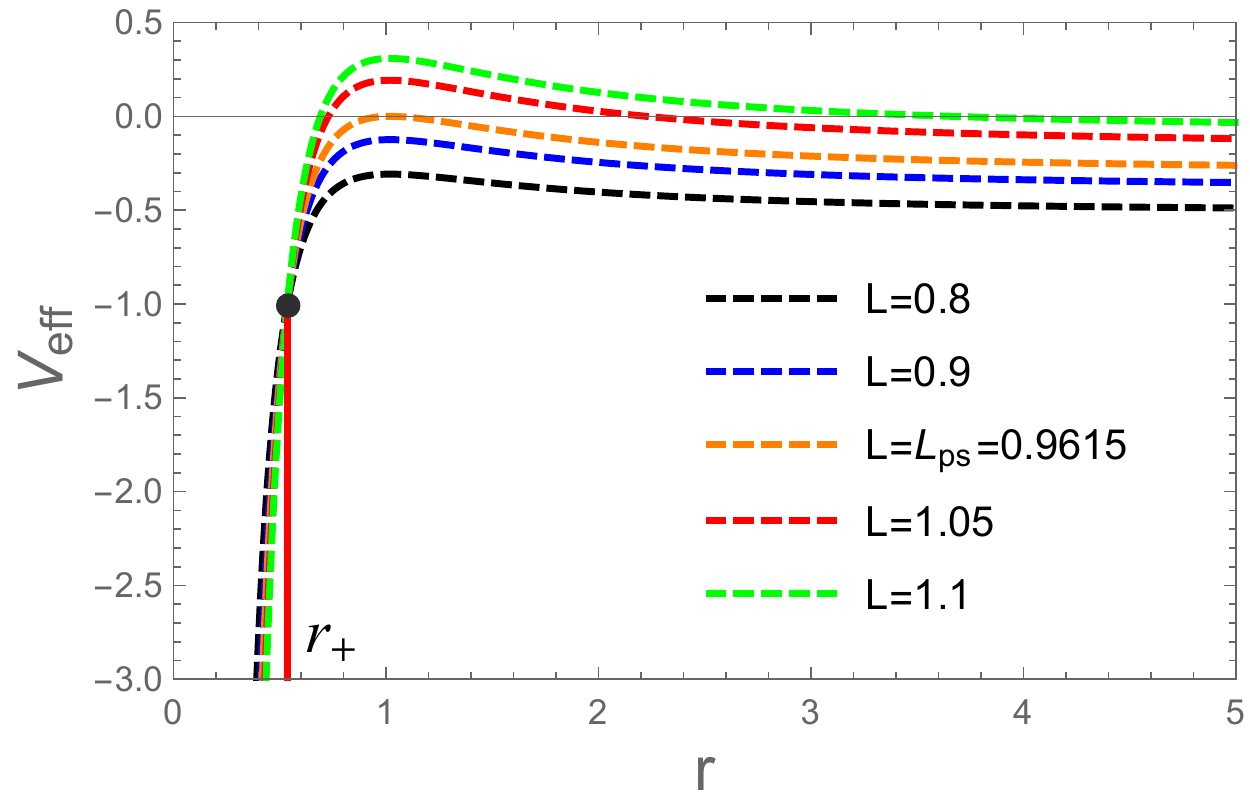}\\
  \caption{(colour online) Portrait of effective potential $V_{\text{eff}}$ as a function of $r$ for various $L$. $(T, P)$ are fixed as $(0.227,0.09)$ to be close to the critical point. The red solid line denotes the position of the horizon where $V_{\text{eff}}=-1$.}\label{Veff}
\end{figure}

As a function of $r$, the effective potential $V_{\text{eff}}$ depends on parameters $(L, \Lambda, m)$ (or $(L, P, r_+)$ or $(L, P, T)$ equivalently). As a simple example, we show the behaviour of the photon's effective potential ($\delta=0$) with $E=1$ for various $L$ in Fig.~\ref{Veff}. $(P,T)$ are chosen to be close to the critical point $(T_c, P_c)$. From the figure, one can see that for each curve there exists a peak $V_{\text{eff}}^{\text{max}}$ at $r=r_{\text{max}}$ which increases as $L$ is increased. Due to the constraint $\dot{r}^2 \geq 0$, the photon would be confined to a certain range in the $r$-direction where the effective potential satisfies $V_{\text{eff}} \leq 0$. So, an ingoing photon from infinity will fall into the black hole inevitably if $V_{\text{eff}}^{\text{max}}<0$ while bounce back if $V_{\text{eff}}^{\text{max}}>0$. For a critical angular momentum $L=L_{ps}$, $V_{\text{eff}}^{\text{max}}=0$ and an interesting thing happens: the ingoing photon will lose both their radial velocity and acceleration at $r=r_{\text{max}}$ completely and then circles the black hole due to its non-vanishing transverse velocity. Thus, in this case $r=r_{\text{max}}$ is called the photon orbit and henceforth we denote it as $r=r_{ps}$, which is determined by the following conditions
\begin{equation}
V_{\text{eff}}\bigg|_{L=L_{ps},r=r_{ps}}=\partial_r {V_{\text{eff}}}\bigg|_{L=L_{ps},r=r_{ps}}=0.
\end{equation}
Of course this photon orbit is unstable as $\partial^2_r {V_{\text{eff}}}\big|_{L=L_{ps},r=r_{ps}}<0$ (it is a local maximum point of the effective potential). Substituting the expression of the effective potential (\ref{EffectivePotential}), these conditions becomes
\begin{eqnarray}
&&u_{ps} \equiv \frac{L_{ps}}{E}=\frac{rR(r)}{\sqrt{f(r})}\bigg|_{r=r_{ps}},\\
&&2\left(rR'(r)+R(r)\right)f(r)-rR(r)f'(r)\bigg|_{r=r_{ps}}=0,
\end{eqnarray}
where $u_{ps}$ is called the minimum impact parameter or the critical angular momentum. From the second equation, the photon sphere radius $r_{ps}$ can be obtained which depends on values of $(P, r_+)$, then the minimum impact parameter $u_{ps}$ is determined by the first equation. At the critical point $(P=P_c=0.0916, r_+=r_{+c}=0.4245)$, we can get $(r_{ps}=r_{psc}=0.7663, u_{ps}=u_{psc}=0.8724)$. In the following discussions, we set $E=1$ for convenience.

\subsection{Relations between photon orbits and phase transitions}

Now let us explore the relations between the photon orbits and phase transitions.

\begin{figure}[!htbp]
  \centering
  \includegraphics[width=0.40\textwidth]{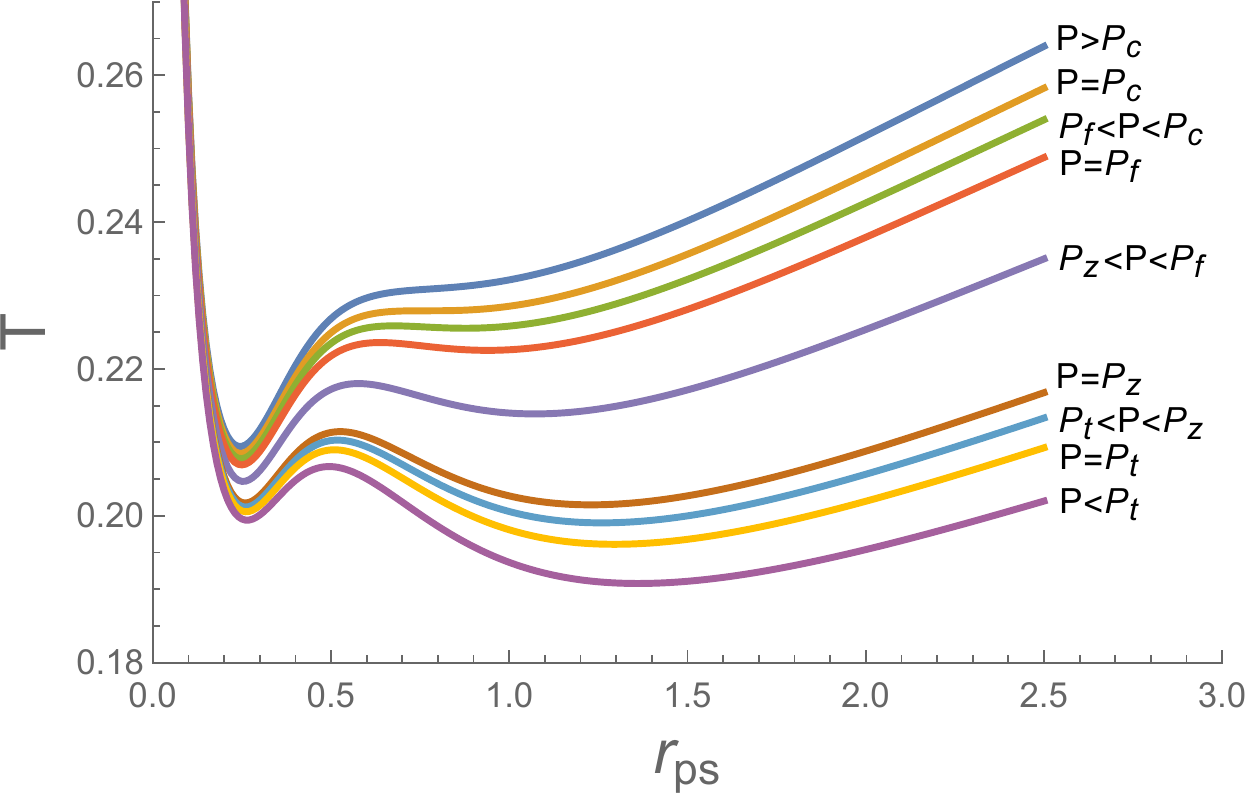}\\
  \caption{(colour online) Hawking temperature versus the photon orbit radius for various pressure. Values of pressure are chosen to be the same as in Fig.~\ref{Trp}.} \label{TrpsFig}
\end{figure}

In Fig.~\ref{TrpsFig}, Hawking temperature $T$ versus the photon orbit radius $r_{ps}$ for various temperature is plotted. From the figure, one can see that isobaric $T-r_{ps}$ curves exhibit similar behaviours as $T-r_+$ curves in Fig~\ref{Trp}. When $P>P_c$, $T-r_{ps}$ curve have only one local minimum point, and thus for a given temperature there only exist one stable branch so that no phase transition can occur; When $P=P_c$, it has one reflection point corresponding to the critical point. When $P<P_c$, two local minimum points appear which means that in a certain range of temperature there are two stable branches, corresponding to LBH and SBH respectively, so that there may be LBH-SBH phase transitions. We denote the two local minimum points as $(r_{ps1},T_{m1})$ and $(r_{ps2},T_{m2})$ with $r_{ps2}>r_{ps1}$. From the figure, one can see that as $P$ is decreased, both $T_{m1}$ and $T_{m2}$ decrease but $T_{m2}$ decreases faster than $T_{m1}$. There exists a marginal value $P_z$ such that for $P<P_z$, $T_{m2}<T_{m1}$ indicating a possible RPT. However, it should be noted that other marginal values $P_f, P_t$ and transition temperatures as well as order of phase transitions can not be read off merely from $T-r_{ps}$ or $T-r_+$ curves. To get these information, we should ask help of the Gibbs free energy.

\begin{figure}[!htbp]
	\centering
	\includegraphics[width=0.40\textwidth]{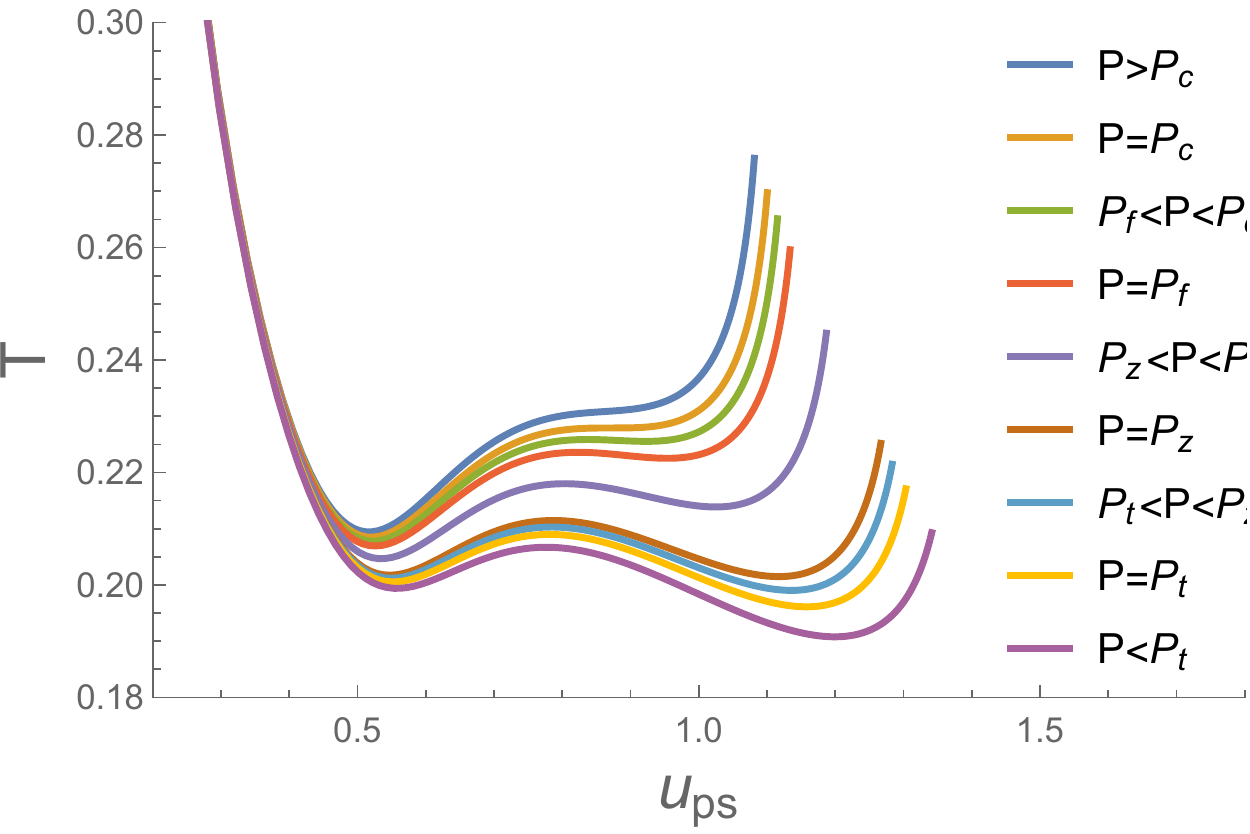}\\
	\caption{(colour online) Hawking temperature versus the minimum impact parameter for various pressure. Values of pressure are chosen to be the same as in Fig.~\ref{Trp}.}\label{TupsFig}
\end{figure}

We also plot the isobaric $T-u_{ps}$ curves in Fig.~\ref{TupsFig}, which show similar behaviours as $T-r_{ps}$ and $T-r_+$ curves.

So, in summary, behaviours of the photon orbit along with the temperature can give us some qualitative information about the phase transitions.

\subsection{Critical behaviour of photon orbits}

From Figs.~\ref{Trp}-\ref{TupsFig}, we know that when $P_t<P<P_c$, there is phase transition between LBH and SBH and the transition may be zeroth-order or first-order or RPT depending on the pressure. Moreover, along the coexisting line, the photon orbit radius $r_{ps}$ will experience a jump. Namely, for a fixed pressure, at the transition temperature, there are two photon orbit radii $r_{ps}^L$ and $r_{ps}^S$ corresponding to LBH and SBH respectively. The situation is the same for the minimum impact parameter.

\begin{figure}[!htbp]
  \centering
  \includegraphics[width=0.5\textwidth]{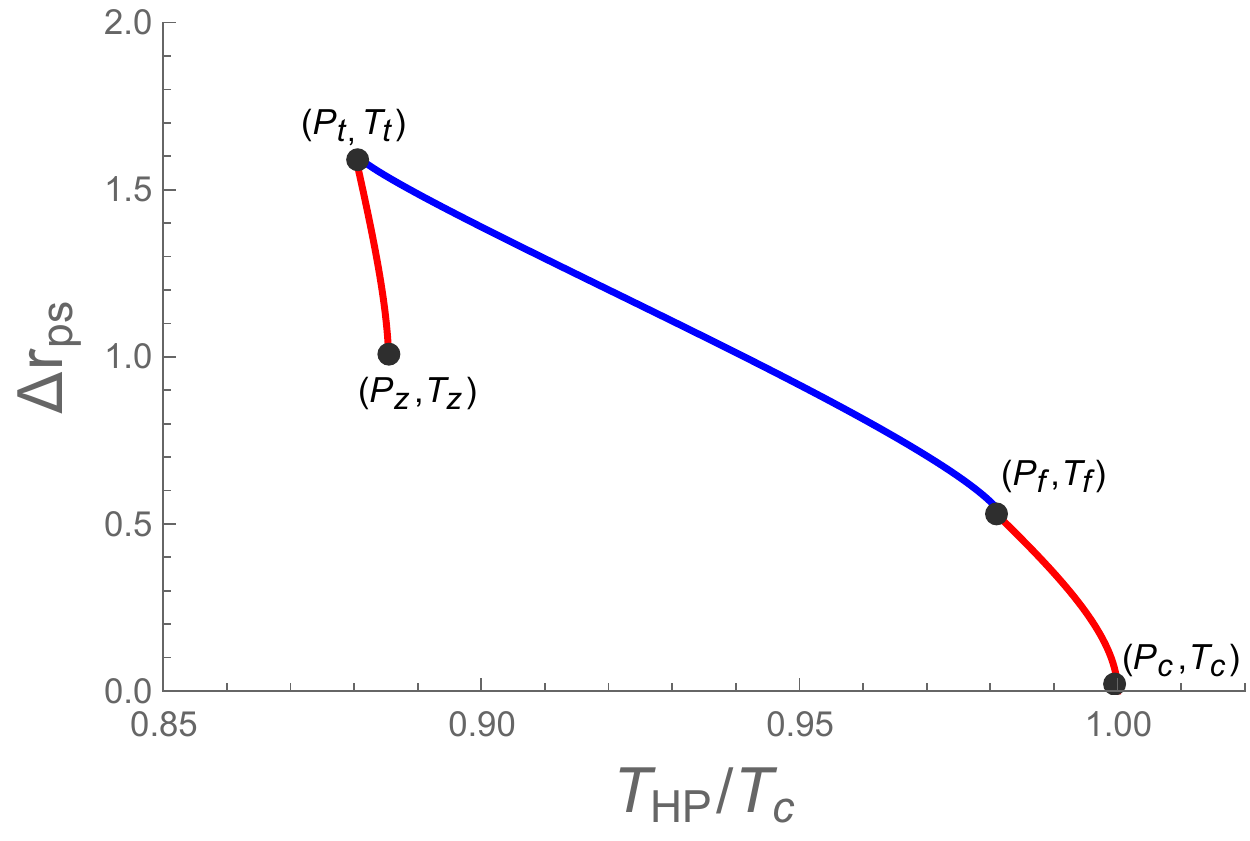}\\
  \caption{Behaviour of $\Delta r_{ps}$ as a function of the transition temperature $T_{HP}$. Marginal points are denoted with black dots. As Fig.~\ref{PTPhaseDiagram}, red solid lines denote zeroth-order coexisting line while blue one denotes first-order coexisting line.}\label{DeltaRpsFig}
\end{figure}

\begin{figure}[!htbp]
	\centering
	\includegraphics[width=0.5\textwidth]{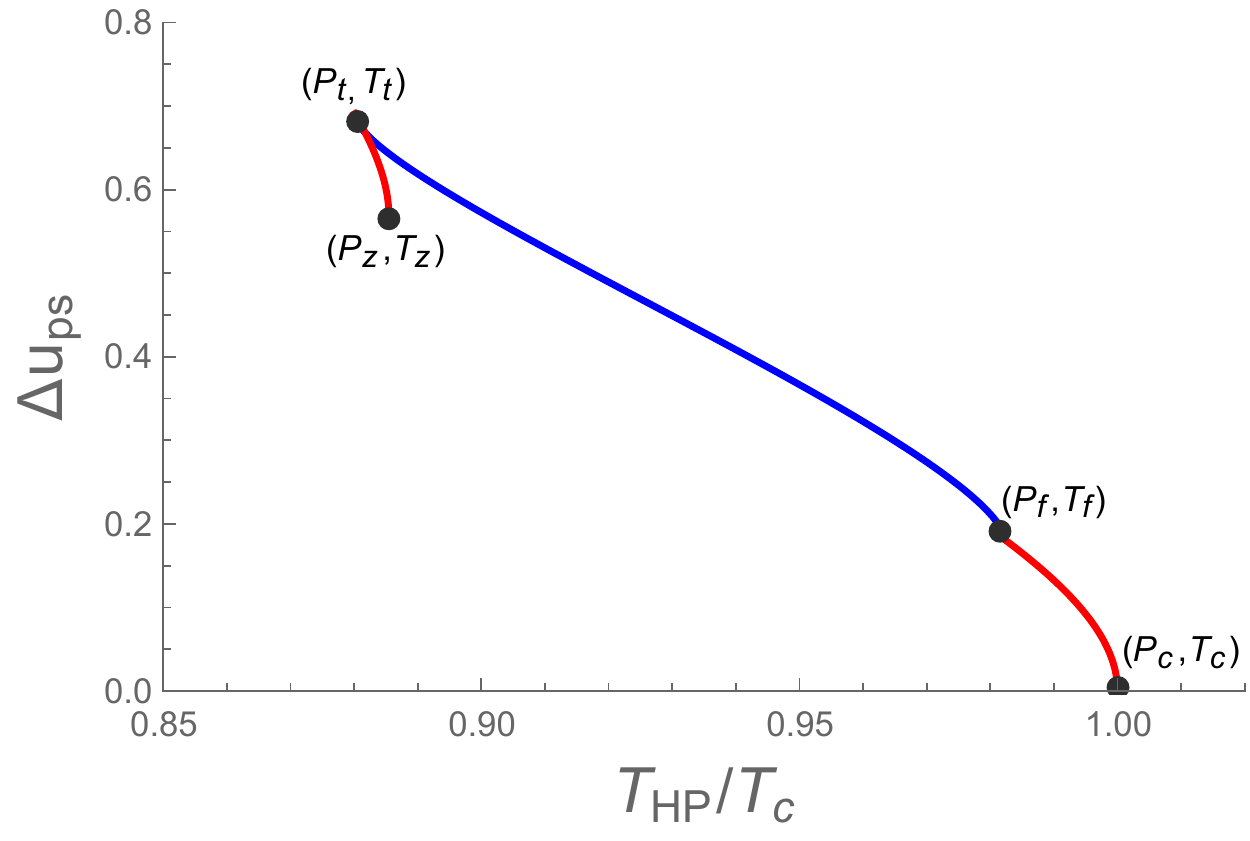}\\
	\caption{Behaviour of $\Delta u_{ps}$ as a function of the transition temperature $T_{HP}$.}\label{DeltaupsFig}
\end{figure}

In Figs.~\ref{DeltaRpsFig} and~\ref{DeltaupsFig}, the transition temperature-dependence of the changes of the photon orbit radius $\Delta r_{ps} \equiv r_{ps}^L-r_{ps}^S$ and the minimum impact parameter $\Delta u_{ps} \equiv u_{ps}^L-u_{ps}^S$ are plotted. From the figures, one can see that $\Delta r_{ps}$ and $\Delta u_{ps}$ have similar behaviours. They both are monotonically decreasing functions of the transition temperature $T_{HP}$ in the range $T_{HP} \in (T_z, T_c)$, while they become  double-valued functions of $T_{HP}$ in the range $T_{HP} \in (T_t, T_z)$ which is reminiscent of RPT. At the critical point, they both vanish. These results imply $r_{ps}$ and $u_{ps}$ may be used as order parameters to describe phase transitions including the RPT.

It is interesting to explore the critical exponents of $\Delta r_{ps}$ and $\Delta u_{ps}$. Following Refs.~\cite{Wei:2017mwc}, by fitting the numerical data with function
\begin{equation}
	\Delta r_{ps}, \Delta u_{ps} \sim a \times \left(1-T_{HP}/T_c\right)^\delta,\label{Fitting}
\end{equation}
it is found that $(a \approx 3.6374, \delta \approx 0.5208)$ for $\Delta r_{ps}$ and $(a \approx 1.3242, \delta \approx 0.5047)$ for $\Delta u_{ps}$. So, the critical exponent $\delta$ is found to be universal and around $\frac{1}{2}$.

\section{Conclusion}\label{partfour}

In this work, in the BIDAdS black hole, we explore the relations between photon orbits and various types of phase transitions. The main results are: (1) As shown in Figs. \ref{TrpsFig} and \ref{TupsFig}, $T-r_{ps}$ and $T-u_{ps}$ exhibit non-monotonic behaviors which are similar to $T-r_+$, thus signalling the existence of LBH-SBH phase transitions. In particular, the marginal value of pressure $P_z$ under which RPT occur can be read off from the behavior of either $T-r_{ps}$ or $T-u_{ps}$ curves. (2) As shown in Figs. \ref{DeltaRpsFig} and \ref{DeltaupsFig}, along the coexisting lines, there will be changes of $r_{ps}$ and $u_{ps}$. Moreover, the two changes are both monotonically decreasing function of the transition temperature $T_{HP}$ in the range $T_{HP} \in (T_z, T_c)$, while they become  double-valued functions of $T_{HP}$ in the range $T_{HP} \in (T_t, T_z)$ which is reminiscent of RPT. And at the critical point, they both vanish. (3) By fitting the changes of $r_{ps}$ and $u_{ps}$ near the critical point (See Eq. (\ref{Fitting})), the critical exponents are found to be universal and around $\frac{1}{2}$. Combing these results, we confirm the conclusion of Refs. \cite{Wei:2017mwc,Wei:2018aqm} that there is a close relation between photon orbits and phase transitions, and the photon orbit radius and the minimum impact parameter can be used as order parameters to describe BH phase transitions. Nevertheless, we should point out zeroth-order and first-order phase transitions can not be distinguished with photon orbits alone.

\section*{Acknowledgements}

This work is supported by National Natural Science Foundation of China (No. 11605155) and partly by the Fundamental Research Funds for the Provincial Universities of Zhejiang in China (No. RF-A2019015). 

\hspace{.5mm}

Declaration: The authors declare that there is no conflict of interest regarding the publication of this paper.

\hspace{.5mm}

Data Availability Statement: The numerical data used to support the findings of this study are included within the article.


\begin{thebibliography}{99}


%\cite{Hawking:1974sw}
\bibitem{Hawking:1974sw}
S.~W.~Hawking,
{\em Particle Creation by Black Holes},
Commun.\ Math.\ Phys.\  {\bf 43}, 199 (1975)
Erratum: [Commun.\ Math.\ Phys.\  {\bf 46}, 206 (1976)].
% doi:10.1007/BF02345020, 10.1007/BF01608497
%%CITATION = doi:10.1007/BF02345020, 10.1007/BF01608497;%%
%7410 citations counted in INSPIRE as of 24 Dec 2018


%\cite{Bekenstein:1973ur}
\bibitem{Bekenstein:1973ur}
J.~D.~Bekenstein,
{\em Black holes and entropy},
Phys.\ Rev.\ D {\bf 7}, 2333 (1973).
% doi:10.1103/PhysRevD.7.2333
%%CITATION = doi:10.1103/PhysRevD.7.2333;%%
%4188 citations counted in INSPIRE as of 24 Dec 2018

%\cite{Maldacena:1997re}
\bibitem{Maldacena:1997re}
J.~M.~Maldacena,
{\em The Large N limit of superconformal field theories and supergravity},
Int.\ J.\ Theor.\ Phys.\  {\bf 38}, 1113 (1999)
[Adv.\ Theor.\ Math.\ Phys.\  {\bf 2}, 231 (1998)]
[hep-th/9711200].
%%CITATION = HEP-TH/9711200;%%
%10255 citations counted in INSPIRE as of 12 Nov 2014

%\cite{Gubser:1998bc}
\bibitem{Gubser:1998bc}
S.~S.~Gubser, I.~R.~Klebanov and A.~M.~Polyakov,
{\em Gauge theory correlators from noncritical string theory},
Phys.\ Lett.\ B {\bf 428}, 105 (1998)
[hep-th/9802109].
%%CITATION = HEP-TH/9802109;%%
%5972 citations counted in INSPIRE as of 12 Nov 2014

%\cite{Witten:1998qj}
\bibitem{Witten:1998qj}
E.~Witten,
{\em Anti-de Sitter space and holography},
Adv.\ Theor.\ Math.\ Phys.\  {\bf 2}, 253 (1998)
[hep-th/9802150].

%\cite{Bardeen:1973gs}
\bibitem{Bardeen:1973gs}
J.~M.~Bardeen, B.~Carter and S.~W.~Hawking,
{\em The Four laws of black hole mechanics},
Commun.\ Math.\ Phys.\  {\bf 31}, 161 (1973).
% doi:10.1007/BF01645742
%%CITATION = doi:10.1007/BF01645742;%%
%1864 citations counted in INSPIRE as of 24 Dec 2018

%\cite{Hawking:1982dh}
\bibitem{Hawking:1982dh}
S.~W.~Hawking and D.~N.~Page,
{\em Thermodynamics of Black Holes in anti-De Sitter Space},
Commun.\ Math.\ Phys.\  {\bf 87}, 577 (1983).
%  doi:10.1007/BF01208266
%%CITATION = doi:10.1007/BF01208266;%%
%1506 citations counted in INSPIRE as of 24 Dec 2018

%\cite{Kubiznak:2014zwa}
\bibitem{Kubiznak:2014zwa}
D.~Kubiznak and R.~B.~Mann,
{\em Black hole chemistry},
Can.\ J.\ Phys.\  {\bf 93}, no. 9, 999 (2015)
%doi:10.1139/cjp-2014-0465
[arXiv:1404.2126 [gr-qc]].
%%CITATION = doi:10.1139/cjp-2014-0465;%%
%70 citations counted in INSPIRE as of 24 Dec 2018

%\cite{Chamblin:1999tk}
\bibitem{Chamblin:1999tk}
A.~Chamblin, R.~Emparan, C.~V.~Johnson and R.~C.~Myers,
{\em Charged AdS black holes and catastrophic holography},
Phys.\ Rev.\ D {\bf 60}, 064018 (1999)
%doi:10.1103/PhysRevD.60.064018
[hep-th/9902170].
%%CITATION = doi:10.1103/PhysRevD.60.064018;%%
%784 citations counted in INSPIRE as of 24 Dec 2018


%\cite{Chamblin:1999hg}
\bibitem{Chamblin:1999hg}
A.~Chamblin, R.~Emparan, C.~V.~Johnson and R.~C.~Myers,
{\em Holography, thermodynamics and fluctuations of charged AdS black holes},
Phys.\ Rev.\ D {\bf 60}, 104026 (1999)
%doi:10.1103/PhysRevD.60.104026
[hep-th/9904197].
%%CITATION = doi:10.1103/PhysRevD.60.104026;%%
%449 citations counted in INSPIRE as of 24 Dec 2018

%\cite{Dolan:2011xt}
\bibitem{Dolan:2011xt}
B.~P.~Dolan,
{\em Pressure and volume in the first law of black hole thermodynamics},
Class.\ Quant.\ Grav.\  {\bf 28}, 235017 (2011)
%doi:10.1088/0264-9381/28/23/235017
[arXiv:1106.6260 [gr-qc]].
%%CITATION = doi:10.1088/0264-9381/28/23/235017;%%
%238 citations counted in INSPIRE as of 24 Dec 2018


%\cite{Kubiznak:2012wp}
\bibitem{Kubiznak:2012wp}
D.~Kubiznak and R.~B.~Mann,
{\em P-V criticality of charged AdS black holes},
JHEP {\bf 1207}, 033 (2012)
%doi:10.1007/JHEP07(2012)033
[arXiv:1205.0559 [hep-th]].
%%CITATION = doi:10.1007/JHEP07(2012)033;%%
%398 citations counted in INSPIRE as of 24 Dec 2018


%\cite{Altamirano:2013uqa}
\bibitem{Altamirano:2013uqa} 
N.~Altamirano, D.~Kubizňák, R.~B.~Mann and Z.~Sherkatghanad,
{\em Kerr-AdS analogue of triple point and solid/liquid/gas phase transition},
Class.\ Quant.\ Grav.\  {\bf 31}, 042001 (2014)
%doi:10.1088/0264-9381/31/4/042001
[arXiv:1308.2672 [hep-th]].
%%CITATION = doi:10.1088/0264-9381/31/4/042001;%%
%129 citations counted in INSPIRE as of 15 Aug 2019

%\cite{Gunasekaran:2012dq}
\bibitem{Gunasekaran:2012dq}
S.~Gunasekaran, R.~B.~Mann and D.~Kubiznak,
{\em Extended phase space thermodynamics for charged and rotating black holes and Born-Infeld vacuum polarization},
JHEP {\bf 1211}, 110 (2012)
%doi:10.1007/JHEP11(2012)110
[arXiv:1208.6251 [hep-th]].
%%CITATION = doi:10.1007/JHEP11(2012)110;%%
%245 citations counted in INSPIRE as of 24 Dec 2018


%\cite{Altamirano:2013ane}
\bibitem{Altamirano:2013ane}
N.~Altamirano, D.~Kubiznak and R.~B.~Mann,
{\em Reentrant phase transitions in rotating anti-de Sitter black holes},
Phys.\ Rev.\ D {\bf 88}, no. 10, 101502 (2013)
%doi:10.1103/PhysRevD.88.101502
[arXiv:1306.5756 [hep-th]].
%%CITATION = doi:10.1103/PhysRevD.88.101502;%%
%149 citations counted in INSPIRE as of 24 Dec 2018

%\cite{Johnson:2014yja}
\bibitem{Johnson:2014yja} 
C.~V.~Johnson,
{\em Holographic Heat Engines},
Class.\ Quant.\ Grav.\  {\bf 31}, 205002 (2014)
%doi:10.1088/0264-9381/31/20/205002
[arXiv:1404.5982 [hep-th]].
%%CITATION = doi:10.1088/0264-9381/31/20/205002;%%
%154 citations counted in INSPIRE as of 15 Aug 2019

%\cite{Kastor:2009wy}
\bibitem{Kastor:2009wy}
D.~Kastor, S.~Ray and J.~Traschen,
{\em Enthalpy and the Mechanics of AdS Black Holes},
Class.\ Quant.\ Grav.\  {\bf 26}, 195011 (2009)
%  doi:10.1088/0264-9381/26/19/195011
[arXiv:0904.2765 [hep-th]].
%%CITATION = doi:10.1088/0264-9381/26/19/195011;%%
%425 citations counted in INSPIRE as of 09 Jan 2019





%\cite{Mann:2016trh}
\bibitem{Mann:2016trh}
R.~B.~Mann,
{\em The Chemistry of Black Holes},
Springer Proc.\ Phys.\  {\bf 170}, 197 (2016).
% doi:10.1007/978-3-319-20046-0_23
%%CITATION = doi:10.1007/978-3-319-20046-0_23;%%
%7 citations counted in INSPIRE as of 25 Dec 2018


%\cite{Kubiznak:2016qmn}
\bibitem{Kubiznak:2016qmn}
D.~Kubiznak, R.~B.~Mann and M.~Teo,
{\em Black hole chemistry: thermodynamics with Lambda},
Class.\ Quant.\ Grav.\  {\bf 34}, no. 6, 063001 (2017)
%doi:10.1088/1361-6382/aa5c69
[arXiv:1608.06147 [hep-th]].
%%CITATION = doi:10.1088/1361-6382/aa5c69;%%
%120 citations counted in INSPIRE as of 24 Dec 2018




%\cite{Jing:2008an}
\bibitem{Jing:2008an} 
J.~Jing and Q.~Pan,
{\em Quasinormal modes and second order thermodynamic phase transition for Reissner-Nordstrom black hole},
Phys.\ Lett.\ B {\bf 660}, 13 (2008)
%doi:10.1016/j.physletb.2007.11.039
[arXiv:0802.0043 [gr-qc]].
%%CITATION = doi:10.1016/j.physletb.2007.11.039;%%
%31 citations counted in INSPIRE as of 12 Aug 2019

%\cite{Berti:2008xu}
\bibitem{Berti:2008xu} 
E.~Berti and V.~Cardoso,
{\em Quasinormal modes and thermodynamic phase transitions},
Phys.\ Rev.\ D {\bf 77}, 087501 (2008)
%doi:10.1103/PhysRevD.77.087501
[arXiv:0802.1889 [hep-th]].
%%CITATION = doi:10.1103/PhysRevD.77.087501;%%
%20 citations counted in INSPIRE as of 12 Aug 2019

%\cite{He:2008im}
\bibitem{He:2008im} 
X.~He, B.~Wang, S.~Chen, R.~G.~Cai and C.~Y.~Lin,
{\em Quasinormal modes in the background of charged Kaluza-Klein black hole with squashed horizons},
Phys.\ Lett.\ B {\bf 665}, 392 (2008)
%doi:10.1016/j.physletb.2008.06.038
[arXiv:0802.2449 [hep-th]].
%%CITATION = doi:10.1016/j.physletb.2008.06.038;%%
%30 citations counted in INSPIRE as of 12 Aug 2019

%\cite{Koutsoumbas:2006xj}
\bibitem{Koutsoumbas:2006xj} 
G.~Koutsoumbas, S.~Musiri, E.~Papantonopoulos and G.~Siopsis,
{\em Quasi-normal Modes of Electromagnetic Perturbations of Four-Dimensional Topological Black Holes with Scalar Hair},
JHEP {\bf 0610}, 006 (2006)
%doi:10.1088/1126-6708/2006/10/006
[hep-th/0606096].
%%CITATION = doi:10.1088/1126-6708/2006/10/006;%%
%55 citations counted in INSPIRE as of 12 Aug 2019

%\cite{Koutsoumbas:2008pw}
\bibitem{Koutsoumbas:2008pw} 
G.~Koutsoumbas, E.~Papantonopoulos and G.~Siopsis,
{\em Phase Transitions in Charged Topological-AdS Black Holes},
JHEP {\bf 0805}, 107 (2008)
%doi:10.1088/1126-6708/2008/05/107
[arXiv:0801.4921 [hep-th]].
%%CITATION = doi:10.1088/1126-6708/2008/05/107;%%
%27 citations counted in INSPIRE as of 12 Aug 2019

%\cite{Shen:2007xk}
\bibitem{Shen:2007xk} 
J.~Shen, B.~Wang, C.~Y.~Lin, R.~G.~Cai and R.~K.~Su,
{\em The phase transition and the Quasi-Normal Modes of black Holes},
JHEP {\bf 0707}, 037 (2007)
%doi:10.1088/1126-6708/2007/07/037
[hep-th/0703102 [HEP-TH]].
%%CITATION = doi:10.1088/1126-6708/2007/07/037;%%
%34 citations counted in INSPIRE as of 12 Aug 2019

%\cite{Rao:2007zzb}
\bibitem{Rao:2007zzb} 
X.~P.~Rao, B.~Wang and G.~H.~Yang,
{\em Quasinormal modes and phase transition of black holes},
Phys.\ Lett.\ B {\bf 649}, 472 (2007)
%doi:10.1016/j.physletb.2007.04.049
[arXiv:0712.0645 [gr-qc]].
%%CITATION = doi:10.1016/j.physletb.2007.04.049;%%
%33 citations counted in INSPIRE as of 12 Aug 2019

%\cite{Myung:2008ze}
\bibitem{Myung:2008ze} 
Y.~S.~Myung,
{\em Phase transition for black holes with scalar hair and topological black holes},
Phys.\ Lett.\ B {\bf 663}, 111 (2008)
%doi:10.1016/j.physletb.2008.03.046
[arXiv:0801.2434 [hep-th]].
%%CITATION = doi:10.1016/j.physletb.2008.03.046;%%
%29 citations counted in INSPIRE as of 12 Aug 2019

%\cite{Liu:2014gvf}
\bibitem{Liu:2014gvf} 
Y.~Liu, D.~C.~Zou and B.~Wang,
{\em Signature of the Van der Waals like small-large charged AdS black hole phase transition in quasinormal modes},
JHEP {\bf 1409}, 179 (2014)
%doi:10.1007/JHEP09(2014)179
[arXiv:1405.2644 [hep-th]].
%%CITATION = doi:10.1007/JHEP09(2014)179;%%
%64 citations counted in INSPIRE as of 12 Aug 2019

%\cite{Mahapatra:2016dae}
\bibitem{Mahapatra:2016dae} 
S.~Mahapatra,
{\em Thermodynamics, Phase Transition and Quasinormal modes with Weyl corrections},
JHEP {\bf 1604}, 142 (2016)
%doi:10.1007/JHEP04(2016)142
[arXiv:1602.03007 [hep-th]].
%%CITATION = doi:10.1007/JHEP04(2016)142;%%
%30 citations counted in INSPIRE as of 12 Aug 2019

%\cite{Chabab:2016cem}
\bibitem{Chabab:2016cem} 
M.~Chabab, H.~El Moumni, S.~Iraoui and K.~Masmar,
{\em Behavior of quasinormal modes and high dimension RN–AdS black hole phase transition},
Eur.\ Phys.\ J.\ C {\bf 76}, no. 12, 676 (2016)
%doi:10.1140/epjc/s10052-016-4518-6
[arXiv:1606.08524 [hep-th]].
%%CITATION = doi:10.1140/epjc/s10052-016-4518-6;%%
%22 citations counted in INSPIRE as of 12 Aug 2019

%\cite{Prasia:2016esx}
\bibitem{Prasia:2016esx} 
P.~Prasia and V.~C.~Kuriakose,
{\em Quasinormal modes and thermodynamics of linearly charged BTZ black holes in massive gravity in (anti) de Sitter space-time},
Eur.\ Phys.\ J.\ C {\bf 77}, no. 1, 27 (2017)
%doi:10.1140/epjc/s10052-016-4591-x
[arXiv:1608.05299 [gr-qc]].
%%CITATION = doi:10.1140/epjc/s10052-016-4591-x;%%
%9 citations counted in INSPIRE as of 12 Aug 2019

%\cite{Zou:2017juz}
\bibitem{Zou:2017juz} 
D.~C.~Zou, Y.~Liu and R.~H.~Yue,
{\em Behavior of quasinormal modes and Van der Waals-like phase transition of charged AdS black holes in massive gravity},
Eur.\ Phys.\ J.\ C {\bf 77}, no. 6, 365 (2017)
%doi:10.1140/epjc/s10052-017-4937-z
[arXiv:1702.08118 [gr-qc]].
%%CITATION = doi:10.1140/epjc/s10052-017-4937-z;%%
%31 citations counted in INSPIRE as of 12 Aug 2019

%\cite{Li:2017kkj}
\bibitem{Li:2017kkj} 
A.~c.~Li, H.~q.~Shi and D.~f.~Zeng,
{\em Phase structure and quasinormal modes of a charged AdS dilaton black hole},
Phys.\ Rev.\ D {\bf 97}, no. 2, 026014 (2018)
%doi:10.1103/PhysRevD.97.026014
[arXiv:1711.04613 [hep-th]].
%%CITATION = doi:10.1103/PhysRevD.97.026014;%%
%5 citations counted in INSPIRE as of 12 Aug 2019

%\cite{Liang:2017ceh}
\bibitem{Liang:2017ceh} 
B.~Liang, S.~W.~Wei and Y.~X.~Liu,
{\em Quasinormal Modes and Van der Waals like phase transition of charged AdS black holes in Lorentz symmetry breaking massive gravity},
Int.\ J.\ Mod.\ Phys.\ D {\bf 28}, no. 09, 1950113 (2019)
%doi:10.1142/S021827181950113X
[arXiv:1712.01545 [gr-qc]].
%%CITATION = doi:10.1142/S021827181950113X;%%
%3 citations counted in INSPIRE as of 12 Aug 2019

\bibitem{Goebel:1972}
C.~G.~Goebel,
{\em Comments on the "vibrations" of a black hole},
Astrophys.\ J.\ {\bf 172}, L95 (1972).

%\cite{Ferrari:1984zz}
\bibitem{Ferrari:1984zz} 
V.~Ferrari and B.~Mashhoon,
{\em New approach to the quasinormal modes of a black hole},
Phys.\ Rev.\ D {\bf 30}, 295 (1984).
%doi:10.1103/PhysRevD.30.295
%%CITATION = doi:10.1103/PhysRevD.30.295;%%
%249 citations counted in INSPIRE as of 12 Aug 2019

%\cite{Mashhoon:1985cya}
\bibitem{Mashhoon:1985cya} 
B.~Mashhoon,
{\em Stability of charged rotating black holes in the eikonal approximation},
Phys.\ Rev.\ D {\bf 31}, no. 2, 290 (1985).
%doi:10.1103/PhysRevD.31.290
%%CITATION = doi:10.1103/PhysRevD.31.290;%%
%88 citations counted in INSPIRE as of 12 Aug 2019

%\cite{Bozza:2002zj}
\bibitem{Bozza:2002zj} 
V.~Bozza,
{\em Gravitational lensing in the strong field limit},
Phys.\ Rev.\ D {\bf 66}, 103001 (2002)
%doi:10.1103/PhysRevD.66.103001
[gr-qc/0208075].
%%CITATION = doi:10.1103/PhysRevD.66.103001;%%
%248 citations counted in INSPIRE as of 15 Aug 2019

%\cite{Cardoso:2008bp}
\bibitem{Cardoso:2008bp} 
V.~Cardoso, A.~S.~Miranda, E.~Berti, H.~Witek and V.~T.~Zanchin,
{\em Geodesic stability, Lyapunov exponents and quasinormal modes},
Phys.\ Rev.\ D {\bf 79}, 064016 (2009)
%doi:10.1103/PhysRevD.79.064016
[arXiv:0812.1806 [hep-th]].
%%CITATION = doi:10.1103/PhysRevD.79.064016;%%
%254 citations counted in INSPIRE as of 15 Aug 2019

%\cite{Stefanov:2010xz}
\bibitem{Stefanov:2010xz} 
I.~Z.~Stefanov, S.~S.~Yazadjiev and G.~G.~Gyulchev,
{\em Connection between Black-Hole Quasinormal Modes and Lensing in the Strong Deflection Limit},
Phys.\ Rev.\ Lett.\  {\bf 104}, 251103 (2010)
%doi:10.1103/PhysRevLett.104.251103
[arXiv:1003.1609 [gr-qc]].
%%CITATION = doi:10.1103/PhysRevLett.104.251103;%%
%43 citations counted in INSPIRE as of 15 Aug 2019

%\cite{Wei:2011zw}
\bibitem{Wei:2011zw} 
S.~W.~Wei, Y.~X.~Liu and H.~Guo,
{\em Relationship between High-Energy Absorption Cross Section and Strong Gravitational Lensing for Black Hole},
Phys.\ Rev.\ D {\bf 84}, 041501 (2011)
%doi:10.1103/PhysRevD.84.041501
[arXiv:1103.3822 [hep-th]].
%%CITATION = doi:10.1103/PhysRevD.84.041501;%%
%20 citations counted in INSPIRE as of 15 Aug 2019

%\cite{Hod:2012nk}
\bibitem{Hod:2012nk} 
S.~Hod,
{\em The fastest way to circle a black hole},
Phys.\ Rev.\ D {\bf 84}, 104024 (2011)
%doi:10.1103/PhysRevD.84.104024
[arXiv:1201.0068 [gr-qc]].
%%CITATION = doi:10.1103/PhysRevD.84.104024;%%
%27 citations counted in INSPIRE as of 15 Aug 2019

%\cite{Wei:2013mda}
\bibitem{Wei:2013mda} 
S.~W.~Wei and Y.~X.~Liu,
{\em Establishing a universal relation between gravitational waves and black hole lensing},
Phys.\ Rev.\ D {\bf 89}, no. 4, 047502 (2014)
%doi:10.1103/PhysRevD.89.047502
[arXiv:1309.6375 [gr-qc]].
%%CITATION = doi:10.1103/PhysRevD.89.047502;%%
%13 citations counted in INSPIRE as of 15 Aug 2019

%\cite{Raffaelli:2014ola}
\bibitem{Raffaelli:2014ola} 
B.~Raffaelli,
{\em Strong gravitational lensing and black hole quasinormal modes: Towards a semiclassical unified description},
Gen.\ Rel.\ Grav.\  {\bf 48}, no. 2, 16 (2016)
%doi:10.1007/s10714-016-2016-7
[arXiv:1412.7333 [gr-qc]].
%%CITATION = doi:10.1007/s10714-016-2016-7;%%
%8 citations counted in INSPIRE as of 15 Aug 2019

%\cite{Cardoso:2016rao}
\bibitem{Cardoso:2016rao} 
V.~Cardoso, E.~Franzin and P.~Pani,
{\em Is the gravitational-wave ringdown a probe of the event horizon?},
Phys.\ Rev.\ Lett.\  {\bf 116}, no. 17, 171101 (2016)
Erratum: [Phys.\ Rev.\ Lett.\  {\bf 117}, no. 8, 089902 (2016)]
%doi:10.1103/PhysRevLett.117.089902, 10.1103/PhysRevLett.116.171101
[arXiv:1602.07309 [gr-qc]].
%%CITATION = doi:10.1103/PhysRevLett.117.089902, 10.1103/PhysRevLett.116.171101;%%
%231 citations counted in INSPIRE as of 15 Aug 2019


%\cite{Konoplya:2017wot}
\bibitem{Konoplya:2017wot} 
R.~A.~Konoplya and Z.~Stuchlík,
{\em Are eikonal quasinormal modes linked to the unstable circular null geodesics?},
Phys.\ Lett.\ B {\bf 771}, 597 (2017)
%doi:10.1016/j.physletb.2017.06.015
[arXiv:1705.05928 [gr-qc]].
%%CITATION = doi:10.1016/j.physletb.2017.06.015;%%
%47 citations counted in INSPIRE as of 15 Aug 2019

%\cite{Dolan:2009nk}
\bibitem{Dolan:2009nk} 
S.~R.~Dolan and A.~C.~Ottewill,
{\em On an Expansion Method for Black Hole Quasinormal Modes and Regge Poles},
Class.\ Quant.\ Grav.\  {\bf 26}, 225003 (2009)
%doi:10.1088/0264-9381/26/22/225003
[arXiv:0908.0329 [gr-qc]].
%%CITATION = doi:10.1088/0264-9381/26/22/225003;%%
%42 citations counted in INSPIRE as of 12 Aug 2019


%\cite{Cvetic:2016bxi}
\bibitem{Cvetic:2016bxi} 
M.~Cvetic, G.~W.~Gibbons and C.~N.~Pope,
{\em Photon Spheres and Sonic Horizons in Black Holes from Supergravity and Other Theories},
Phys.\ Rev.\ D {\bf 94}, no. 10, 106005 (2016)
%doi:10.1103/PhysRevD.94.106005
[arXiv:1608.02202 [gr-qc]].
%%CITATION = doi:10.1103/PhysRevD.94.106005;%%
%17 citations counted in INSPIRE as of 15 Aug 2019

%\cite{Gibbons:2008ru}
\bibitem{Gibbons:2008ru} 
G.~W.~Gibbons, C.~M.~Warnick and M.~C.~Werner,
{\em Light-bending in Schwarzschild-de-Sitter: Projective geometry of the optical metric},
Class.\ Quant.\ Grav.\  {\bf 25}, 245009 (2008)
%doi:10.1088/0264-9381/25/24/245009
[arXiv:0808.3074 [gr-qc]].
%%CITATION = doi:10.1088/0264-9381/25/24/245009;%%
%33 citations counted in INSPIRE as of 15 Aug 2019

%\cite{Wei:2017mwc}
\bibitem{Wei:2017mwc} 
S.~W.~Wei and Y.~X.~Liu,
{\em Photon orbits and thermodynamic phase transition of $d$-dimensional charged AdS black holes},
Phys.\ Rev.\ D {\bf 97}, no. 10, 104027 (2018)
%doi:10.1103/PhysRevD.97.104027
[arXiv:1711.01522 [gr-qc]].
%%CITATION = doi:10.1103/PhysRevD.97.104027;%%
%10 citations counted in INSPIRE as of 15 Aug 2019

%\cite{Wei:2018aqm}
\bibitem{Wei:2018aqm} 
S.~W.~Wei, Y.~X.~Liu and Y.~Q.~Wang,
{\em Probing the relationship between the null geodesics and thermodynamic phase transition for rotating Kerr-AdS black holes},
Phys.\ Rev.\ D {\bf 99}, no. 4, 044013 (2019)
%doi:10.1103/PhysRevD.99.044013
[arXiv:1807.03455 [gr-qc]].
%%CITATION = doi:10.1103/PhysRevD.99.044013;%%
%5 citations counted in INSPIRE as of 15 Aug 2019

%\cite{Han:2018ooi}
\bibitem{Han:2018ooi} 
S.~Z.~Han, J.~Jiang, M.~Zhang and W.~B.~Liu,
{\em Photon orbits and thermodynamic phase transition in Gauss-Bonnet AdS black holes},
arXiv:1812.11862 [gr-qc].
%%CITATION = ARXIV:1812.11862;%%
%1 citations counted in INSPIRE as of 15 Aug 2019

%\cite{Chabab:2019kfs}
\bibitem{Chabab:2019kfs} 
M.~Chabab, H.~El Moumni, S.~Iraoui and K.~Masmar,
{\em Probing correlation between photon orbits and phase structure of charged AdS black hole in massive gravity background},
arXiv:1902.00557 [hep-th].
%%CITATION = ARXIV:1902.00557;%%
%1 citations counted in INSPIRE as of 15 Aug 2019

%\cite{Xu:2019yub}
\bibitem{Xu:2019yub} 
Y.~M.~Xu, H.~M.~Wang, Y.~X.~Liu and S.~W.~Wei,
{\em Photon sphere and reentrant phase transition of charged Born-Infeld-AdS black holes},
arXiv:1906.03334 [gr-qc].
%%CITATION = ARXIV:1906.03334;%%

%\cite{Zhang:2019tzi}
\bibitem{Zhang:2019tzi} 
M.~Zhang, S.~Z.~Han, J.~Jiang and W.~B.~Liu,
{\em Circular orbit of a test particle and phase transition of a black hole},
Phys.\ Rev.\ D {\bf 99}, no. 6, 065016 (2019)
%doi:10.1103/PhysRevD.99.065016
[arXiv:1903.08293 [hep-th]].
%%CITATION = doi:10.1103/PhysRevD.99.065016;%%
%2 citations counted in INSPIRE as of 15 Aug 2019




%\cite{Dehyadegari:2017flm}
\bibitem{Dehyadegari:2017flm}
A.~Dehyadegari, A.~Sheykhi and A.~Montakhab,
{\em Novel phase transition in charged dilaton black holes},
Phys.\ Rev.\ D {\bf 96}, no. 8, 084012 (2017)
%doi:10.1103/PhysRevD.96.084012
[arXiv:1707.05307 [hep-th]].
%%CITATION = doi:10.1103/PhysRevD.96.084012;%%
%6 citations counted in INSPIRE as of 24 Dec 2018

%\cite{Momennia:2017hsc}
\bibitem{Momennia:2017hsc}
S.~H.~Hendi and M.~Momennia,
{\em Reentrant phase transition of Born-Infeld-dilaton black holes},
Eur.\ Phys.\ J.\ C {\bf 78}, no. 10, 800 (2018)
%doi:10.1140/epjc/s10052-018-6278-y
[arXiv:1709.09039 [gr-qc]].
%%CITATION = doi:10.1140/epjc/s10052-018-6278-y;%%
%6 citations counted in INSPIRE as of 24 Dec 2018


%\cite{Dayyani:2017fuz}
\bibitem{Dayyani:2017fuz}
Z.~Dayyani, A.~Sheykhi, M.~H.~Dehghani and S.~Hajkhalili,
{\em Critical behavior and phase transition of dilaton black holes with nonlinear electrodynamics},
Eur.\ Phys.\ J.\ C {\bf 78}, no. 2, 152 (2018)
%  doi:10.1140/epjc/s10052-018-5623-5
[arXiv:1709.06875 [gr-qc]].
%%CITATION = doi:10.1140/epjc/s10052-018-5623-5;%%
%8 citations counted in INSPIRE as of 22 Feb 2019

%\cite{Sheykhi:2006dz}
\bibitem{Sheykhi:2006dz}
A.~Sheykhi and N.~Riazi,
{\em Thermodynamics of black holes in (n+1)-dimensional Einstein-Born-Infeld dilaton gravity},
Phys.\ Rev.\ D {\bf 75}, 024021 (2007)
%  doi:10.1103/PhysRevD.75.024021
[hep-th/0610085].
%%CITATION = doi:10.1103/PhysRevD.75.024021;%%
%59 citations counted in INSPIRE as of 22 Feb 2019


%\cite{Sheykhi:2007gw}
\bibitem{Sheykhi:2007gw}
A.~Sheykhi,
{\em Topological Born-Infeld-dilaton black holes},
Phys.\ Lett.\ B {\bf 662}, 7 (2008)
%doi:10.1016/j.physletb.2008.02.017
[arXiv:0710.3827 [hep-th]].
%%CITATION = doi:10.1016/j.physletb.2008.02.017;%%
%49 citations counted in INSPIRE as of 24 Dec 2018

%\cite{Chen:2018icg}
\bibitem{Chen:2018icg} 
Y.~Chen, H.~Li and S.~J.~Zhang,
{\em Microscopic explanation for black hole phase transitions via Ruppeiner geometry: two competing mechanisms},
arXiv:1812.11765 [hep-th].
%%CITATION = ARXIV:1812.11765;%%



	
	
\end{thebibliography}
\end{document}